\documentclass[a4paper]{article}
\usepackage[all]{xy}\usepackage[latin1]{inputenc}        
\usepackage[dvips]{graphics,graphicx}
\usepackage{amsfonts,amssymb,amsmath,color,mathrsfs, amstext}
\usepackage{amsbsy, amsopn, amscd, amsxtra, amsthm,authblk}
\usepackage{enumerate,algorithmicx,algorithm}
\usepackage{algpseudocode}
\usepackage{upref}
\usepackage{geometry}
\usepackage{float}
\usepackage{yhmath}
\usepackage{booktabs}
\usepackage{bm}
\usepackage{multirow}
\usepackage{makecell}
\usepackage[colorlinks,
            linkcolor=red,
            anchorcolor=red,
            citecolor=red
            ]{hyperref}

\numberwithin{equation}{section}

\def\b{\boldsymbol}

\DeclareMathOperator{\erf}{erf}
\DeclareMathOperator{\erfc}{erfc}

\usepackage{blindtext}

\newtheorem{remark}{Remark}

\begin{document}

\title{A note on accurate pressure calculations of Coulomb systems with periodic boundary conditions}
\author[1,2]{Lei Li \thanks{leili2010@sjtu.edu.cn}}
	\author[1]{Jiuyang Liang \thanks{liangjiuyang@sjtu.edu.cn}}
	\author[1,2]{Zhenli Xu \thanks{xuzl@sjtu.edu.cn}}
	\affil[1]{School of Mathematical Sciences, Shanghai Jiao Tong University, Shanghai, 200240, P. R. China}
	\affil[2]{Institute of Natural Sciences, MOE-LSC and CMA-Shanghai, Shanghai Jiao Tong University, Shanghai, 200240, P. R. China}

\date{}

\maketitle

\begin{abstract}
	In this note, we address some issues concerning the accurate pressure calculation of Coulomb systems with periodic boundary conditions. First, we prove that the formulas for the excess part of the pressure with Ewald summation also reduce to the ensemble average of one-third of the ratio between the potential energy and the volume so that the comments on our previous work in 
 a recent paper by [Onegin~\emph{et al}.,~J. Phys. A: Math.~Theor.~57 (2024) 205002] are incorrect.
 Second, we demonstrate that in charge non-neutral systems, the pressure expression must be corrected to include interactions with the neutralizing background. This addresses the issues about pressure computation in LAMMPS raised in the paper by Onegin {\it et al.}. Numerical experiments are performed to verify that the pressure obtained via Ewald summation with corrected terms agrees with the average pressure using thermodynamics for the non-neutral OCP system, and are independent of the splitting parameter in the Ewald summation. 
\end{abstract}

\section{Introduction}
Pressure is a fundamental thermodynamic quantity. For systems with short-range interactions, the pressure calculation is generally straightforward using the standard virial expression~\cite{frenkel2001understanding,Thompson2009JCP}. However, challenges arise when dealing with long-range interactions and fully/quasi-periodic boundary conditions. In molecular dynamics (MD) simulations, Coulomb interactions are typically addressed through Ewald-type methods~\cite{Hockney1988Computer,Darden1993JCP,Jin2021SISC} where the long-range part is calculated in the Fourier space. The Ewald-type methods significantly reduce the complexity of energy and force calculations, but they result in expressions that are no longer in pairwise form, making the virial expression inapplicable. To overcome this issue, a thermodynamic definition of pressure is introduced for periodic Coulomb systems, relying on the differentiation of the Helmholtz free energy with respect to volume~\cite{brown1995general,Hummer1998JCP,sega2016pressure,Liang_2022}. When the energy is evaluated using Ewald summation, both the real-space and Fourier-space contributions explicitly depend on the volume and contribute to the pressure~\cite{brown1995general}. This approach has been widely adopted in mainstream MD software packages such as LAMMPS~\cite{thompson2021lammps} and GROMACS~\cite{berendsen1995gromacs}. 

However, debate on this issue remains ongoing. In a recent study~\cite{Onegin_2024}, it was claimed that the splitting parameter of the Ewald summation as a volume-related variable could significantly influence the thermodynamic pressure, which reported a significant gap between the thermodynamic pressure and the virial pressure by simulations of a one-component plasma (OCP) system using LAMMPS software. These findings seem non-physical, as pressure should not depend on mathematical techniques of algorithms. In this note, we first prove that the formulas for the excess part of the pressure with Ewald summation in our previous work agrees with the ensemble average of one third of the ratio between potential energy and the volume for the Coulomb system. 
We further identify that the strange phenomena observed in~\cite{Onegin_2024} arise from the lack of pressure contributions related to background charges in a non-neutral system. Although the uniform background does not affect the forces~\cite{Ballenegger2008JCP,Ballenegger2009JCP}, the corresponding energy does depend on volume. We then derive the appropriate correction terms. Numerical results verify that the modified pressure is independent of the splitting parameter in the Ewald summation. These corrections, which prevent significant incorrect cell scaling, are especially crucial when using a large real-space cutoff. Our results can be seamlessly extended to infinite boundary terms and quasi-2D systems with surface charges~\cite{yi2017note,gan2024fast}.

The organization of the remainder of this note is as follows: Section 2 reviews the derivation of thermodynamic and virial pressures. In Section 3, we study the pressure in systems with long-range Coulomb interactions and derive the corrections for non-neutral systems. Section 4 presents the results of numerical tests. Finally, Section 5 provides concluding remarks.

\section{The formulas for virial and thermodynamic pressure}


We consider a cubic cell $\Omega$ (assume isotropic cell fluctuation) of volume $V=L^3$ containing $N$ charged particles. The Cartesian coordinates and momenta of these particles are denoted by $\{\bm{r}_1,\cdots,\bm{r}_N\}$ and $\{\bm{p}_1,\cdots,\bm{p}_N\}$, respectively. Let us denote that $d\bm{r}=d\bm{r}_1\cdots d\bm{r}_N$ and $d\bm{p}=d\bm{p}_1\cdots d\bm{p}_N$. The derivation starts with the pressure of system from statistical physics
\begin{gather}\label{eq:pressure1}
	\widetilde{P}=\frac{1}{\beta}\left(\frac{\partial \log Q}{\partial V}\right)\bigg{|}_T,
\end{gather}
where $T$ is the temperature, $\beta=(k_{\text{B}}T)^{-1}$ is the inverse thermal energy with $k_{\text{B}}$ the Boltzmann constant, and
\begin{gather}
	Q(N, V, T)=\int \exp\left[-\beta\left(\sum_{i=1}^{N} \frac{|\b{p}_i|^2}{2m_i}+U(\{\b{r}_i\}; V)\right)\right]d\b{r} d\b{p}
\end{gather}
denotes the partition function for the NVT ensemble. To compute the derivative on $V$, the position and momentum vectors are scaled as 
\begin{equation}
\bm{r}_i=V^{1/3} \bm{s}_i, \quad \bm{p}_i=V^{-1/3}\bm{p}_i^{s}
\end{equation}
in~\cite{Liang_2022}, where the Jacobian for the transform $(\bm{r}_i,\bm{p}_i)\rightarrow(\bm{s}_i,\bm{p}_i^s)$ is $1$. With the change of variables, the dependence on the volume is then moved to the integrand totally. By using this technique, the pressure formula is derived
\begin{equation}\label{eq:pressure}
\widetilde{P}=\left\langle- \frac{\partial (K+U)}{\partial V}{\bigg|}_{\bm{s}_i,\bm{p}_i^s} \right \rangle
=\left\langle \frac{1}{3V}\sum_i \frac{|\b{p}_i|^2}{m_i}-\frac{\partial U(\{V^{1/3}\b{s}_i\}, V)}{\partial V}\bigg{|}_{\b{s}_i\leftarrow V^{-1/3}\b{r}_i} \right\rangle,
\end{equation}
where $K(V,\bm{p}_i^s):=V^{-2/3}\sum_{i}|\bm{p}^s_i|^2/2m_i$ denotes the kinetic energy and $\langle\cdot\rangle$ indicates the ensemble average under $Q(N,V,T)$, which would be the expression for the pressure we use for the Coulomb system.

Now, let us make some discussions on the comments about Liang {\it et al.}~\cite{Liang_2022} in a recent paper by Onegin {\it et al.}~\cite{Onegin_2024}. This paper did not consider the momentum variable explicitly, but considered the partition function
\begin{equation}
Q=\int \exp(-\beta U)d\bm{r},
\end{equation}
and applied the same formula for pressure 
\begin{equation}
P_F=\frac{1}{\beta}\frac{\partial \log Q}{\partial V}{\bigg|}_T.
\end{equation}
To find the derivative, they also performed scaling $V=\gamma^3 V_0$ and $\bm{r}_i=\gamma \bm{r}_{i,0}$ so that $\bm{r}_{i,0}$ are the coordinates in the cell of the initial volume $V_0$ and the dependence on $V$ will not appear on the integration domain. 
The difference is that in their scaling, $\gamma$ is dimensionless while in our scaling $s_i$ is dimensionless. Neverthless, mathematically, the $\gamma$ parameter in their scaling plays the similar role as our $L$. They then derived that (see Eq.~(12) in Ref.~\cite{Onegin_2024} where they used $\mathcal{P}_F=\beta P_FV/N$)
\begin{equation}\label{eq::virial}
P_F=\frac{N}{\beta V}-\left\langle \frac{\partial U}{\partial V}{\bigg|}_T \right\rangle.
\end{equation}
They mentioned that the first term is due to the kinetic part (or ``ideal gas contribution'' as they called) while the second one is due to the interaction. The first term appears here because they only consider the $\b{r}_i$ variables instead of the conjugate variables $(\bm{r}_i,\bm{p}_i)$ as we did. In this sense, the term $N/\beta V$ corresponds to 
$\left\langle (1/3V) \sum_i  |\b{p}_i|^2/m_i \right\rangle$ in our formula.

Hence, at this stage, except the contribution from the kinetic part, both the formulas in Refs.~\cite{Onegin_2024} and~\cite{Liang_2022} for the excess component of pressure are actually given by the same formula
\begin{equation}\label{eq::Pexcess}
P_{\text{excess}}=\left\langle -\frac{\partial U(\{V^{1/3}\b{s}_i\}, V)}{\partial V}\bigg{|}_{\b{s}_i\leftarrow V^{-1/3}\b{r}_i} \right\rangle
\end{equation}
or $-\left\langle \partial U/\partial V |_T \right\rangle$ in their notation. 
In their formula, $\partial U/\partial V |_T$ is computed by fixing $V_0, \bm{r}_{i,0}$, etc., so this derivative on $V$ is essentially the same. 

Starting from Eq.~\eqref{eq::Pexcess} and applying the chain rule, one can compute that (see Eq.~(3.8) in Ref.~\cite{Liang_2022})
\begin{gather}
	\label{eq}P_{\text{excess}} =\left\langle -\frac{1}{3V} \bm{r}_i \cdot \nabla_{\bm{r}_i} U({\bm{r}_i}; V) - \frac{\partial }{\partial V} U({\bm{r}_i}; V) \right\rangle.
\end{gather}
In Ref.~\cite{Onegin_2024}, they presented the so-called ``corrected virial pressure'' where the corresponding excess part is (Eqs.~(16) and (18) in Ref.~\cite{Onegin_2024})
\begin{gather}
	P_{\text{excess}} = -\left\langle \frac{\partial U}{\partial V} {\bigg|}_T \right\rangle = \left\langle\frac{1}{3V} \sum_{i=1}^{N} \bm{r}_i \cdot \bm{f}_i - \frac{L}{3V} \frac{\partial U}{\partial L} {\bigg|}_{T, \bm{r}_i} \right\rangle,
\end{gather}
where $\bm{f}_i = -\nabla_{\bm{r}_i} U$ is the force on the $i$th particle. Clearly, the formula in Eq.~\eqref{eq} matches theirs. However, they incorrectly commented in Ref.~\cite{Onegin_2024} (penultimate paragraph of Page 11):
\begin{center}``Thus, instead of calculating $(\partial U / \partial L)_{T,\bm{r}_i}$ in Eq. (16) (or Eq. (3.8) in Ref.~[35]), the authors evaluate $(\partial U / \partial L)_{T, \bm{s}_i}$. This error leads to an incorrect pressure contribution $\cdots$''\end{center}
Here, the mentioned ``Ref.~[35]'' is the previous work \cite{Liang_2022} of the authors.
In Eq.~\eqref{eq::Pexcess}, the partial derivative is taken by fixing $\bm{s}_i$, but after applying the chain rule, in the last term of Eq.~\eqref{eq} above (or Eq. (3.8) of~\cite{Liang_2022}), the partial derivative is clearly evaluated by fixing $\bm{r}_i$. We believe that the authors of Ref.~\cite{Onegin_2024} misunderstood the mathematical formulations in \cite{Liang_2022}.

In summary, the ``corrected virial pressure'' in Ref.~\cite{Onegin_2024} and previous works Refs.~\cite{brown1995general,Liang_2022} are, in fact, the same. Moreover, as noted in~\cite{Liang_2022}, using the corrected virial pressure Eq.~\eqref{eq} for Coulomb systems involving Ewald summation is inefficient. Instead, one should use Eq.~\eqref{eq::Pexcess} directly. This will be further clarified in the next section.

\section{Pressure calculation in Coulomb systems}

In \cite{Liang_2022}, the authors rigorously derived the pressure for Coulomb systems using the Ewald summation. However, the authors of Ref.~\cite{Onegin_2024} claimed that our derivation in Eqs.~(4.10)-(4.13) of~\cite{Liang_2022} does not satisfy the relation $P_{\mathrm{excess}}=\langle U/3V\rangle$ and thus contains errors. Additionally, they argued that the pressure formulas used in LAMMPS~\cite{thompson2021lammps}, which have been employed for decades, are also incorrect based on their numerical simulations of an one-component plasma system. In this section, we 
prove that the formulas in our previous work also satisfy the relation $P_{\mathrm{excess}}=\langle U/3V\rangle$ and provide the correct formulas for the charge non-neutral system so that their comments are incorrect for both our work and the LAMMPS software.

\subsection{Pressure formulas under the Ewald summation}
By using the homogeneity of the Coulomb potential, $U(\{\gamma \bm{r}_i\}; \gamma L)=\gamma^{-1}U(\{\bm{r}_i\}; L)$, the authors of \cite{Onegin_2024} derived (see Eq.~(34) in their paper~\cite{Onegin_2024}, using $P_F$ instead of $\mathcal{P}_F$)
\begin{gather}\label{eq:pressureusingU}
	P_F=\frac{N}{\beta V}+\left\langle \frac{ U}{3V} \right\rangle.
\end{gather}
This is correct, and in our notation, it corresponds to
\begin{gather}\label{eq:derivativerequirement}
	-\frac{\partial U(\{V^{1/3}\b{s}_i\}, V)}{\partial V}\bigg{|}_{\b{s}_i\leftarrow V^{-1/3}\b{r}_i}
	= \frac{U}{3V}.
\end{gather}
Having verified that \eqref{eq:pressureusingU} (Eq. (34) in their paper~\cite{Onegin_2024}) matches the corrected virial pressure (Eq. (18) in~\cite{Onegin_2024}), they commented on Page 11 of Ref.~\cite{Onegin_2024}: 
\begin{center}
	``Eq.~(4.13) in Ref. [35] for the pressure fails to reduce to the simple form (34). Although the dependence of the potential on the cell length is taken into account in [35], Eq. (4.13) turns out to be incorrect.''
\end{center} 
This remark is invalid. Their concern is that when using the Ewald splitting technique~\cite{ewald1921berechnung} 
\begin{equation}\label{eq::ew}
	\frac{1}{r}=\frac{\erf(\sqrt{\alpha} r)}{r}+\frac{\erfc(\sqrt{\alpha})}{r},
\end{equation}
the requirement \eqref{eq:derivativerequirement} is violated. Here, $\erf(x):=\frac{2}{\sqrt{\pi}}\int_{0}^xe^{-u^2}du$ and $\erfc(x):=1-\erf(x)$ represent the error function and complementary error function, respectively, and $\alpha$ is a positive parameter.  They also commented on Page 11 of Ref.~\cite{Onegin_2024}:
\begin{center}
 	``Although equation (42) is correct, the quantity $\sqrt{\alpha}$ has an inverse length dimension, since $r\propto L$, $\alpha \propto 1/L^2$. $\cdots$ Thus, the derivative of $\alpha$ over the volume should be taken into account, which is not done during the derivation of equations (4.10)-(4.13) in [35].''
\end{center}
However, Eq.~\eqref{eq::ew} is an \emph{exact} mathematical equality regardless of the choice of $\alpha$ and whether the Fourier spectral expansion is applied. Thus, the comments concerning the derivative of $\alpha$ are entirely baseless.

For clarification, we prove \eqref{eq:derivativerequirement} directly for the pressure formulas in~\cite{Liang_2022}. Assume that the charge neutrality condition
\begin{equation}\label{eq::chargeneutrality}
	\sum_{i=1}^{N}q_i=0
\end{equation}
is satisfied. In the Ewald summation, the Coulomb energy is decomposed into two parts, $U:=U_1+U_2$, where 
\begin{gather}\label{eq:U1fourier}
	U_1=\frac{2\pi}{V} \sum_{\b{k}\neq \b{0}}\frac{e^{-|\b{k}|^2/(4\alpha)}}{|\b{k}|^2} |\rho(\b{k})|^2-\sqrt{\frac{\alpha}{\pi}}\sum_{i=1}^{N} q_i^2,
\end{gather}
with $\b{k}=2\pi V^{-1/3}\b{m}$ for $\b{m}\in\mathbb{Z}^3$
and $\rho(\b{k}):=\sum_i q_i e^{i \b{k}\cdot \b{r}_i}$, and 
\begin{equation}\label{U2}
	U_2=\frac{1}{2}\sum_{\b{n}\in\mathbb{Z}^3}{'}\sum_{i,j=1}^{N}  q_i   q_j \frac{1}{|\b{r}_{ij}+L\b{n}|}
	\mathrm{erfc}\left(\sqrt{\alpha}|\b{r}_{ij}+L\b{n}|\right).
\end{equation}
with $\bm{r}_{ij}=\bm{r}_j-\bm{r}_i$. In Eq.~\eqref{U2}, the prime indicates that the case of $i=j$ and $\bm{n}=\bm{0}$ is excluded in the summation. The derivatives of $U_1$ and $U_2$ with respect to $V$ are computed in~\cite{Liang_2022}:
\begin{equation}
	-\frac{\partial U_1}{\partial V}
	=\frac{2\pi}{V^2}\sum_{\b{k}\neq \b{0}}\frac{|\rho(\b{k})|^2}{|\b{k}|^2}e^{-|\b{k}|^2/(4\alpha)}
	\left(\frac{1}{3}-\frac{|\b{k}|^2}{6\alpha}\right),
\end{equation}
and 
\begin{equation}\label{eq:UV}
	-\frac{\partial U_2}{\partial V} = \frac{1}{6V}\sum_{\b{n}\in\mathbb{Z}^3}{'}\sum_{i,j=1}^{N}  q_iq_j G(|\bm{r}_{ij}+L\bm{n}|)|\bm{r}_{ij}+L\bm{n}|,
\end{equation}
where
\begin{equation}
	G(r):=\frac{\erfc(\sqrt{\alpha}r)}{r^2}+\frac{2\sqrt{\alpha}e^{-\alpha r^2}}{\sqrt{ \pi}r}.
\end{equation}
Note that there is a sign typo in (4.11) and (4.12) in Ref.~\cite{Liang_2022} in comparison with Eq.~\eqref{eq:UV}.  Though $- \partial_V U_{\ell} |_{\b{s}_i}
\neq U_{\ell}/(3V)$ for $\ell=1,2$ due to the Ewald splitting, we actually have the following claim:
\begin{equation}\label{prop::3V}
-\frac{\partial U_1}{\partial V}
-\frac{\partial U_2}{\partial V}
=\frac{U_1+U_2}{3V},
\end{equation}
which is actually a reformulation of Eq.~\eqref{eq:derivativerequirement}. It can be verified directly that
\begin{equation}\label{eq::U1}
\frac{U_1}{3V}+\frac{\partial U_1}{\partial V}
=-\frac{\sqrt{\alpha}}{3\sqrt{\pi}V}\sum_{i=1}^{N} q_i^2
+\frac{\pi}{3\alpha V^2}\sum_{\bm{k}\neq \bm{0}}|\rho(\bm{k})|^2e^{-|\bm{k}|^2/(4\alpha)}
\end{equation}
and 
\begin{equation}\label{eq::U2}
\frac{U_2}{3V}+\frac{\partial U_2}{\partial V}
=\frac{\sqrt{\alpha}}{3\sqrt{\pi}V}\sum_i q_i^2-\frac{1}{3V}
\sum_{i=1}^Nq_i \phi(\bm{r}_i),
\end{equation}
where
\begin{equation}
	\phi(\bm{r}):=\sqrt{\frac{\alpha}{\pi}}\sum_{j=1}^{N} q_j\sum_{\bm{n}\in\mathbb{Z}^3}
	e^{-\alpha |\bm{r}_{j}-\bm{r}+L\bm{n}|^2}.
\end{equation}
To prove Eq.~\eqref{prop::3V}, let us consider the Fourier transform of $\phi(\bm{r})$:
\begin{equation}
\begin{split}
\widetilde{\phi}(\bm{k})&=\sqrt{\frac{\alpha}{\pi}}\sum_{j=1}^{N} q_j\int_{\Omega} \sum_{\bm{n}\in\mathbb{Z}^3}
e^{-\alpha |\bm{r}_{j}-\bm{r}+L\bm{n}|^2}e^{-i\bm{k}\cdot \bm{r}}d\bm{r}\\
&=\sqrt{\frac{\alpha}{\pi}}\sum_{j=1}^{N} q_j\int_{\mathbb{R}^3}e^{-\alpha |\bm{r}_j-\bm{r}|^2}
e^{-i\bm{k}\cdot \bm{r}}d\bm{r}\\
&=\frac{\pi}{\alpha}e^{-|\bm{k}|^2/(4\alpha)}\sum_{j=1}^{N} q_j e^{-i\bm{k}\cdot \bm{r}_j},
\end{split}
\end{equation}
where one has $\widetilde{\phi}(\bm{0})=0$ due to the charge neutrality condition. The Fourier spectral expansion of $\phi(\bm{r})$ is then given by
\begin{equation}\label{eq::3.15}
	\phi(\bm{r})=\frac{\pi}{\alpha V}\sum_{\bm{k}\neq\bm{0}}e^{-|\bm{k}|^2/(4\alpha)}\sum_{j=1}^{N} q_j e^{-i\bm{k}\cdot (\bm{r}_j-\bm{r})}.
\end{equation}
Substituting Eq.~\eqref{eq::3.15} into Eq.~\eqref{eq::U2} and combining it with Eq.~\eqref{eq::U1} yields the result in Eq.~\eqref{prop::3V}. This invalidates the remark in Ref.~\cite{Onegin_2024}. Furthermore, their discussion on the derivative of $\alpha$ with respect to $L$ on Page 11 of~\cite{Onegin_2024} is also unjustified. Mathematically, the Ewald decomposition is accurate whether $\alpha$ depends on $L$ or not. For the NPT ensemble, where the volume varies, it is computationally more efficient to fix $\alpha$~\cite{thompson2021lammps}.



In conclusion, the derivation of pressure of~\cite{Liang_2022} is correct, and the criticism in Ref.~\cite{Onegin_2024} is invalid. The correct pressure under the Ewald summation is given by
\begin{gather}\label{eq:pressurecubic}
\begin{split}
	P_{\mathrm{ins}}&=\frac{1}{3V}\sum_{i=1}^{N} \frac{|\b{p}_i|^2}{m_i}+\frac{2\pi}{V^2}\sum_{\b{k}\neq \b{0}}\frac{|\rho(\b{k})|^2}{|\b{k}|^2}e^{-|\b{k}|^2/(4\alpha)}
	\left(\frac{1}{3}-\frac{|\b{k}|^2}{6\alpha}\right)\\
	&\quad+\frac{1}{6V}\sum_{\b{n}\in\mathbb{Z}^3}{'}\sum_{i,j=1}^{N}  q_iq_j G(|\bm{r}_{ij}+L\bm{n}|)|\bm{r}_{ij}+L\bm{n}|\\
	& =:P_{1}+P_{2}+P_3
\end{split}
\end{gather}
as shown in Eq.~(4.13) in Ref.~\cite{Liang_2022}, consistent with \cite{brown1995general,sega2016pressure}. 

\subsection{Pressure correction for non-neutral systems} 
In Section 5 of Ref.~\cite{Onegin_2024}, the authors showed an inconsistency between thermodynamic pressure Eq.~\eqref{eq:pressureusingU} and virial pressure Eq.~\eqref{eq:pressurecubic} obtained via LAMMPS. They argued that this gap is due to the absence of the $\partial U/\partial V$ term and suggested using Eq.~(18) in Ref.~\cite{Onegin_2024}. However, this claim is incorrect. Although some documentation may be outdated, the pressure calculation code in LAMMPS explicitly includes this term. The identified gap is not due to the formula but because the simulated system is \emph{non-neutral} and requires additional pressure correction. 

Below, we derive this correction explicitly. Let us consider a system that some charges (electrons, counterions, etc.) are treated implicitly so that the charge neutrality condition Eq.~\eqref{eq::chargeneutrality} is violated. An example is the OCP systems simulated in \cite{Onegin_2024}. Let us define the total charge by
\begin{equation}
	Q_{\text{tot}}:=\sum_{i=1}^{N}q_i.
\end{equation} 
To eliminate the divergence of Coulomb energy, a general way is to assume a homogeneous and isotropic charge distribution $\rho_{\text{back}}=-Q_{\text{tot}}/V$ which serves as the background. The Ewald summation can be used in such systems, where the total Coulomb energy is modified as~\cite{Ballenegger2008JCP,Ballenegger2009JCP}
\begin{equation}\label{eq::corrU}
	U_{\text{corr}}=U_1+U_2+U_{\text{c-b}}^{\text{corr}}+U_{\text{b-b}}^{\text{corr}}.
\end{equation}
In Eq.~\eqref{eq::corrU}, the third term
\begin{equation}
\begin{split}
U_{\text{c}\text{-}\text{b}}^{\text{corr}}:=&\sum_{i=1}^{N}q_i\int_{\Omega}\sum_{\bm{n}\in\mathbb{Z}^3}\frac{\erfc(\alpha |\bm{r}-\bm{r}_i+L\bm{n}|)}{|\bm{r}-\bm{r}_i+L\bm{n}|}\rho_{\text{back}}d\bm{r}\\
=&\sum_{i=1}^{N}q_i\int_{\mathbb{R}^3}\frac{\erfc(\alpha|\bm{r}|)}{|\bm{r}|}\rho_{\text{back}}d\bm{r}
=-\frac{\pi Q_{\text{tot}}^2}{\alpha^2V}
\end{split}
\end{equation}
represents the interaction between charged particles and the neutralizing background. The last term
\begin{equation}
	U_{\text{b-b}}^{\text{corr}}:=\frac{1}{2}\int_{\Omega}\rho_{\text{back}}\int_{\Omega}\sum_{\bm{n}\in\mathbb{Z}^3} \frac{\erfc(\alpha|\bm{r}-\bm{r}'+\bm{n}L|)}{|\bm{r}-\bm{r}'+\bm{n}L|}\rho_{\text{back}}d\bm{r}'d\bm{r}=\frac{\pi Q_{\text{tot}}^2}{2\alpha^2V}
\end{equation}
is due to background-background interactions. The interaction energies between particles and the background and background-background due to $\erf(r)/r$ have nontrivial contribution only for the $\bm{k}=0$ Fourier mode, which exactly cancels the corresponding singularity in the charge-charge interactions. In other words, these energies are included to cancel the otherwise divergent $\bm{k}=0$ term in the Fourier sum~\cite{Ballenegger2009JCP} so that we do not need to consider them explicitly.

The two energy corrections in \eqref{eq::corrU} also contribute to the virial pressure:
\begin{equation}\label{eq::Pins1}
	P_{\text{corr}}=P_{\text{ins}}+P_{\text{c-b}}^{\text{corr}}+P_{\text{b-b}}^{\text{corr}}
\end{equation}
where the first three terms are given via Eq.~\eqref{eq:pressurecubic},
\begin{equation}
	P_{\text{c-b}}^{\text{corr}}:=-\frac{\partial U_{\text{c-b}}^{\text{corr}}}{\partial V}=-\frac{\pi Q_{\text{tot}}^2}{\alpha^2 V^2},\quad \text{and} \quad P_{\text{b-b}}^{\text{corr}}:=-\frac{\partial U_{\text{b-b}}^{\text{corr}}}{\partial V}=\frac{\pi Q_{\text{tot}}^2}{2\alpha^2 V^2}.
\end{equation}
In Ref.~\cite{Onegin_2024}, the authors incorrectly treated $U_{\text{c-b}}^{\text{corr}}$ and $U_{\text{b-b}}^{\text{corr}}$ as negligible constants, overlooking their contributions to pressure. Specifically, they just used $P_{\text{ins}}$ for pressure so that it coincides with $\langle(U_1+U_2)/3V\rangle$. Additionally, in the LAMMPS code, while $U_{\text{c-b}}^{\text{corr}} $ and $ U_{\text{b-b}}^{\text{corr}} $ are correctly included for energy computation, the corresponding corrections $ P_{\text{c-b}}^{\text{corr}} $ and $P_{\text{b-b}}^{\text{corr}}$ are omitted in the pressure calculation, which is also incorrect. This omission explains the discrepancy between the thermodynamic and virial pressures reported in Ref.~\cite{Onegin_2024}. For accurate NPT ensemble simulations, it is essential to include these corrections to ensure proper scaling of the simulation box.


Similar corrections arise when treating the infinite boundary term. In Eq.~\eqref{eq:U1fourier}, the zero-frequency term in the Fourier component of the energy is neglected. This is equivalent to impose a tin-foil boundary condition~\cite{zhonghan2014JCTC} at $r\rightarrow \infty$. If other infinite boundary conditions are considered, the corresponding corrections should be added. For instance, a spherical infinite boundary leads to the correction 
\begin{equation}
U_{\bm{0}}^{\text{corr}}:=\frac{2\pi}{3V}\left|\sum_{i=1}^{N}q_i\bm{r}_i\right|^2, \quad\text{and}\quad P_{\bm{0}}^{\text{corr}}:=-\frac{\partial U_{\bm{0}}^{\text{corr}}}{\partial V}=\frac{2\pi}{9V^2}\left|\sum_{i=1}^{N}q_i\bm{r}_i\right|^2,
\end{equation}
for the energy and pressure, respectively. Although many previous studies have discussed energy corrections~\cite{zhonghan2014JCTC,Liang2023SISRBSOG}, the corresponding pressure corrections are rarely mentioned. Properly incorporating these corrections aids in accurately calculating pressure-related quantities.

\begin{remark}
	As a side remark, the aforementioned formulas assume that the momenta of all particles are canonically sampled. However, if some degrees of freedom are conserved, the ideal gas component of pressure should be adjusted to $N_{\text{f}}/(3\beta V)$ (see ~\cite{Kolafa2024JCP} for a recent discussion), where
	\begin{equation}
		N_{\text{f}}:=3N-N_{\text{c}}-N_{\text{m}}
	\end{equation}  
	represents the number of degrees of freedom. Here, $N_{\text{c}}$ is the number of independent constraints, and $N_{\text{m}}$ is the number of	integrals of motion excluding total energy.  One has $N_{\text{m}}=3$ in periodic boundary conditions for three conserved total momenta. 
\end{remark}

\section{Numerical Examples}
We will now validate our arguments by some numerical tests. We perform MD simulations on two systems: a non-neutral OCP system, as used in Ref.~\cite{Onegin_2024}, and a neutral monovalent implicit electrolyte solution system, previously simulated in Ref.~\cite{LIANG2022108332}.

\begin{figure}[!ht]	
	\centering
	\includegraphics[width=0.99\textwidth]{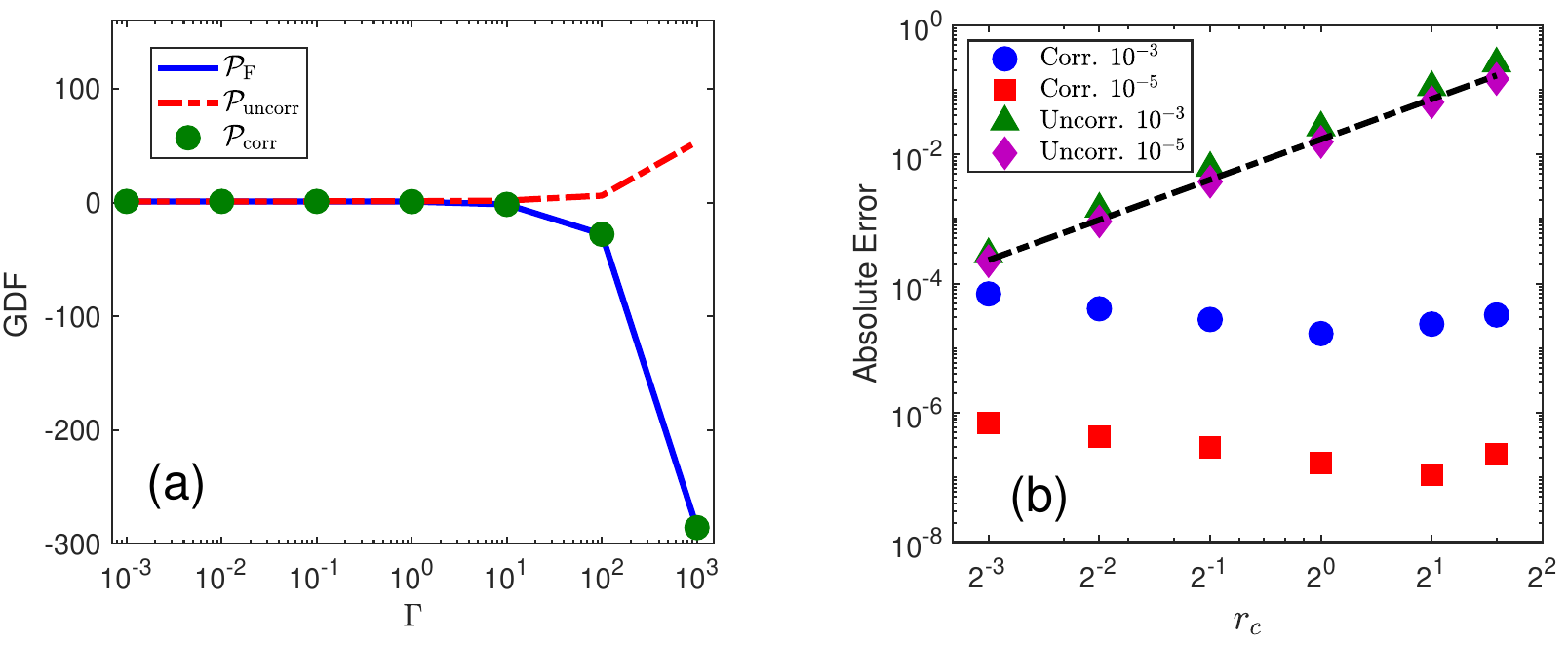}
	\caption{(a): The  gas deviation factors (GDFs) are plotted against the coupling parameter $\Gamma$ for an OCP system. $\mathcal{P}_F$, $\mathcal{P}_{\text{uncorr}}$, and $\mathcal{P}_{\text{corr}}$ represent the GDFs calculated using the thermodynamic, the uncorrected virial, and the corrected pressures, respectively. (b): The absolute error of pressure against real-space cutoff $r_c$. Data are shown for corrected and uncorrected pressures with different Ewald summation precisions. The dash-dotted line in (b) indicates a fitting of $O(r_c^2)$ scaling.}
	\label{fig:conv_result}
\end{figure}

The OCP system consists of $N=100$ particles with strength $+1$. These particles interact through a Coulomb potential evaluated using Ewald summation. The side length is fixed as  $L=7.48\sigma$, where $\sigma$ is the Lennard-Jones (LJ) unit of length. The thermodynamics of the system depends on the coupling parameter $\Gamma=\beta/r_a$, with $r_a=(4\pi N/3)^{-1/3}L$ being the ion-sphere radius. Simulations are conducted in LAMMPS~\cite{thompson2021lammps} under the NVT ensemble, using the input file from the supplementary data of Ref.~\cite{Onegin_2024}. We utilize a time step of $0.001\,\tau$, where $\tau=\sigma\sqrt{m_0/(k_{\text{B}}T)}$ is the unit of time with $m_0$ the ion mass. After discarding the initial non-equilibrium region, the production period consists $10^5$ time steps. To compare with Ref.~\cite{Onegin_2024}, we calculate three gas deviation factors (GDFs) of free charges:
\begin{equation}
	\mathcal{P}_{F}=\frac{\beta V P_F}{N},\quad \mathcal{P}_{\text{uncorr}}=\frac{\beta VP_{\text{ins}}}{N},\quad\text{and}\quad \mathcal{P}_{\text{corr}}=\frac{\beta VP_{\text{corr}}}{N},
\end{equation}
where $P_{F}$ is the thermodynamic pressure from Eq.~\eqref{eq:pressureusingU} with the exact energy $U_{\text{corr}}$. Note that $U_{\text{corr}}$ and $P_{\text{ins}}$ are the default outputs for potential energy and pressure in LAMMPS. Results against the coupling factor $\Gamma$ are shown in Figure~\ref{fig:conv_result}(a), where the real-space cutoff and relative precision for Ewald summation are fixed at $r_c=2\sigma$ and $10^{-4}$, respectively. $\mathcal{P}_{F}$ and $\mathcal{P}_{\text{corr}}$ show good agreement, while the uncorrected GDF $\mathcal{P}_{\text{uncorr}}$ from LAMMPS diverges significantly as $\Gamma$ increases. This gap between thermodynamic and virial pressure in LAMMPS, as observed in previous studies~\cite{Onegin_2024}, is due to nothing but the lack of pressure corrections for non-neutral systems. 

Figure~\ref{fig:conv_result}(b) displays the absolute error of uncorrected and corrected pressure, $P_{\text{ins}}$ and $P_{\text{corr}}$, against different real-space cutoffs $r_c$ in the Ewald summation. The coupling parameter is set as $\Gamma=10^3$. The reference solution uses the $P_{F}$ value calculated with $r_c=1\sigma$ and a precision of $10^{-8}$. The results show that the error between the corrected pressure for non-neutral systems and the exact value is comparable to the truncation error of Ewald summation. In contrast, the uncorrected virial pressure deviates significantly from the exact value by $P_{\text{c-b}}^{\text{corr}}\sim \alpha^{-2}$. The error estimates for the Ewald summation~\cite{Kolafa1992MolSimul} suggest the optimal parameter $\alpha\sim r_c^{-1}$ for a fixed precision, leading to a quadratic increase in error with the cutoff $r_c$. The fitting lines in the figure confirm this relationship.

\begin{figure*}[!ht]	
	\centering
 \includegraphics[width=0.99\textwidth]{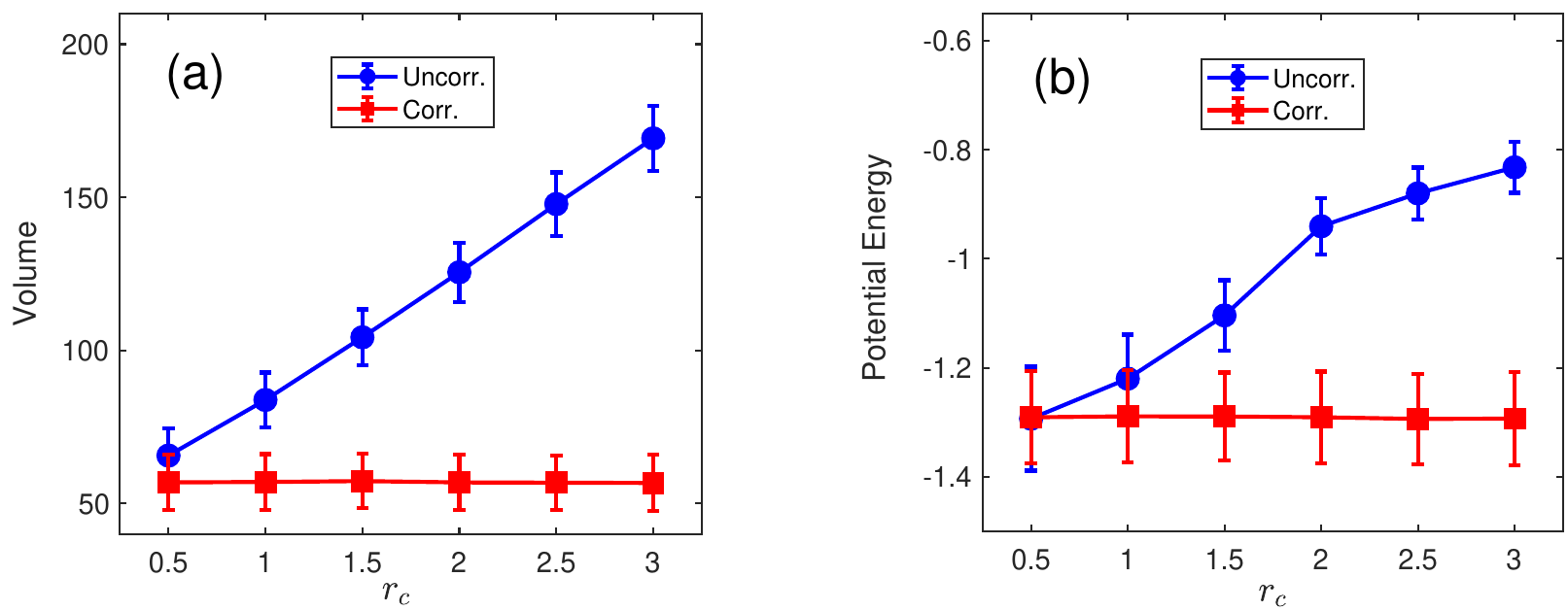}
	\caption{The ensemble average of (a) volume and (b) potential energy under the NPT ensemble. Here, ``Corr.'' indicates that the pressure is computed using the corrected formula Eq.~\eqref{eq::Pins1}. The remaining four bars represent data generated using different cutoffs $r_c$, with pressure evaluated using the uncorrected $P_{\text{ins}}$ (default in LAMMPS). The error bars in these figures show the standard deviation. }
	\label{fig:conv_result_3}
\end{figure*}
\begin{sloppypar}
To assess the impact of neglecting the charge-background contribution $P_{\text{c-b}}^{\text{corr}}$ and background-background contribution $P_{\text{b-b}}^{\text{corr}}$ in the total pressure on the box scaling of non-neutral systems, we conducted simulations using LAMMPS for $10^6$ steps under the NPT ensemble. The equilibrium pressure is set to $1$ (LJ unit), with a coupling coefficient of $\Gamma=1$. In Figure~\ref{fig:conv_result_3}(a-b), we plot the ensemble average of the volume of box and the potential energy of the free particle component. The error level for the Ewald summation is set to $10^{-4}$. It is evident from the results that simulations performed with the uncorrected pressure $P_{\text{uncorr}}$ exhibit significant deviations in equilibrium box scaling and consequent potential energy due to the absence of parts $P_{\text{c-b}}^{\text{corr}}$ and $P_{\text{b-b}}^{\text{corr}}$. Since $r_c\sim \alpha^{-1}$, 
from the expressions of both, it can be observed that increasing $r_c$ leads to a decrease in $\alpha$, thereby resulting in an increasingly significant overestimation of pressure.
\end{sloppypar}
\renewcommand\arraystretch{1.4}
\begin{table}[!htbp]
	\caption{Thermodynamic pressure $P_F$ and virial pressure $P_{\text{ins}}$ calculated by LAMMPS, plotted against different parameter $\alpha$. }
	\centering
	\scalebox{1}{
	\begin{tabular}{c|ccccc}
		\hline & $\alpha=0.1$ & $\alpha=0.2$ & $\alpha=0.4$ & $\alpha=0.8$ & $\alpha=1.6$ \\\hline 
		$P_{F}$ & $0.0019831273$ & $0.0019831273$ &$0.0019831273$ & $ 0.0019831273$ & $0.0019831272$ \\\hline 
		$P_{\text{ins}}$ & $0.0019831269$ & $0.0019831273$ & $ 0.0019831272$ & $0.0019831271$ & $0.0019831273$ \\\hline
		\end{tabular}
	}
	\label{tabl:parameter}
\end{table}

To confirm Eq.~\eqref{prop::3V} for the Ewald summation, we calculate the thermodynamic pressure $P_F$ and the virial pressure $P_{\text{ins}}$ for an electrolyte solution system with a side length $L=100\sigma$, consisting of $2000$ oppositely charged monovalent ions. Since the system is charge neutral, both $P_{\text{c-b}}^{\text{corr}}$ and $P_{\text{b-b}}^{\text{corr}}$ are zero. We perform simulations over $10^6$ steps for averaging. The relative precision of the Ewald summation is set to $10^{-6}$~\cite{Kolafa1992MolSimul}. The results in Table~\ref{tabl:parameter} show a strong agreement between $P_F$ and $P_{\text{ins}}$ and indicate that the pressure is independent of $\alpha$, up to a decomposition error. Thus, the concern about the derivative of $\alpha$~\cite{Onegin_2024} is unnecessary.

\section{Conclusion}

In this note, we have addressed some issues concerning the pressure in systems with Coulomb interactions and periodic boundary conditions using Ewald summation. First, we prove that the formulas for the excess part of pressure with Ewald summation in the previous work~\cite{Liang_2022} of the authors is consistent with the ensemble average of one-third of the ratio between potential energy and volume, so that the critique in Ref.~\cite{Onegin_2024}  is unfounded. Second, we also identified that the observed gap between thermodynamic and virial pressures in OCP systems, as noted in~\cite{Onegin_2024}, results from the absence of non-neutral corrections, and we derived these necessary corrections. These corrections are essential for accurate cell rescaling in NPT simulations. Our findings are supported by numerical examples. Extending these results to non-neutral quasi-2D systems~\cite{yi2017note,gan2024fast} are also straightforward.

\section*{Acknowledgement}
This work is financially supported by the National Key R$\&$D Program of China, Project Number 2021YFA1002800 and the Shanghai Science and Technology Commission (grant No. 21JC1403700). J. L. and Z. X. are partially supported by the Natural Science Foundation of China (grant No. 12325113).

\bibliographystyle{elsart-num.bst}

\end{document}